# Weighted Monte Carlo: Calibrating the Smile and Preserving Martingale Condition

A. Elices[1], E. Giménez[2]

*Abstract*—**Weighted Monte Carlo prices exotic options calibrating the probabilities of previously generated paths by a regular Monte Carlo to fit a set of option premiums. When only vanilla call and put options and forward prices are considered, the Martingale condition might not be preserved. This paper shows that this is indeed the case and overcomes the problem by adding additional synthetic options. A robust, fast and easy-to-implement calibration algorithm is presented. The results are illustrated with a geometric cliquet option which shows how the price impact can be significant.**

## I. INTRODUCTION

Taking into account the smile effect to price exotic derivatives is an issue of major concern for both traders and practitioners as the impact of this effect in price and hedging can be very significant. Under efficient market hypothesis, skew and smile on the vanilla option market can be explained either by the fact that volatility is not constant or because the asset price process might have jumps.

There have been many proposals for asset price processes to capture the above-mentioned effects. [Merton, 1976] deduced the risk neutral partial integro-differential equation that a derivative product should follow under the assumption of a jump diffusion process with constant volatility. He also found the analytical solution for a call option. [Dupire, 1994] proposed a model based on the assumption that the volatility of the asset is a deterministic function of time and price. This model has no general analytical solution for call options.

[Heston, 1993] found the analytical solution for a call option assuming that the volatility follows the continuous version of a GARCH process. Other proposals for stochastic volatility models can be found in [Hull and White, 1987], [Stein, 1991] and [Hagan et al, 2002]. Other authors have focused on providing no-arbitrage pricing models that might not have an explicit expression of the asset price process but are able to replicate a broad set of traded options. The implied tree by [Derman and Kani, 1994] and the weighted Monte Carlo by [Avellaneda et al, 2001] belong to this category.

The conceptual idea of the Weighted Monte Carlo is very simple. An initial set of paths is generated. These paths can incorporate any desired feature of the underlying process such as mean reverting stochastic volatility, different regimes of volatility, volatility correlated with the underlying process or even jumps. For a regular Monte Carlo, all these paths will have the same probability. The Weighted Monte Carlo modifies the probability of each path to replicate the prices of the options of the smile at different maturities. As the number of degrees of freedom is fantastic (as many as the number of paths), it is clear that to fit a few option prices, an infinite number of solutions must exist in the absence of arbitrage. From all those, the solution chosen is such that the posterior distribution is as close as possible to the prior in a certain sense. This solution is found solving an optimisation problem under as many constraints as option prices to fit.

The main drawback of this method is that when several maturities are simulated and the smile effect is steep, the probability distribution at each maturity after calibration may be significantly different from the original one. The method has no control over the resulting price process after calibration if no additional action is taken. In particular, the martingale property may not be satisfied. The main contribution of this paper overcomes this problem by introducing additional constraints which force the conditional martingale condition to be satisfied. This is the only property that all processes must share for consistent pricing. Another contribution of this paper is a robust, fast and easy to implement solution algorithm for the optimisation problem. This algorithm allows handling a considerable number of constraints (the geometric cliquet example of section IV has 215 constraints).

Section II reviews the calibration method and presents a robust, fast and easy-to-implement algorithm to find the solution. Section III presents the formulation of the additional constraints to force the conditional martingale property of the underlying process. Section IV presents an example of a geometric cliquet option which is considerably undervalued when the martingale condition is not enforced (about 160 basis points). This conclusion is very much in agreement with trader experience and market pricing. Section V presents the conclusions of the paper.

[1] Model Validation Group, Area of Methodology, División General de Riesgos, Santander, Ciudad Financiera Santander, Avda. Cantabria s/n, 28660 Boadilla del Monte, Spain, *aelices@gruposantander.com*.

[2] Model Development Group, Front Office, Caixabank, Av. Diagonal, 621-629 T.I. P13, Barcelona, Spain, *eduard.gimenez.f@lacaixa.es*.

## II. REVIEW OF THE METHOD

### A. Formulation of the problem

Pricing an european option $\Pi$ using Monte Carlo involves the simulation of $v$ underlying paths according to the hypothesis considered, the calculation of the discounted payoffs at maturity ($g_i$) and the calculation of the price according to equation (1).

$$\Pi_g = \frac{1}{v}\sum_{i=1}^{v} g_i \qquad (1)$$

The prices of the call and put options at different strikes and maturities will in general not be reproduced from the paths generated. The weighted Monte Carlo produces a set of probabilities $p_i$ close to $q_i = 1/v$ in a certain sense providing a new price for $\Pi$ as shown in equation (2).

$$\Pi_g = \sum_{i=1}^{v} g_i p_i \qquad (2)$$

The "distance" between two distributions $\mathbf{p}$ and $\mathbf{q}$ is defined using the relative entropy (although strictly speaking the relative entropy is not a distance) as defined in equation (3):

$$D(\mathbf{p}/\mathbf{q}) = \sum_{i=1}^{v} p_i \ln\left(\frac{p_i}{q_i}\right) \qquad (3)$$

When the probabilities of the prior distribution are equal, equation (3) reduces to equation (4).

$$D(\mathbf{p}/\mathbf{q}) = \ln v + \sum_{i=1}^{v} p_i \ln p_i \qquad (4)$$

Equation (5) presents a matrix of discounted payoffs for the $N$ securities to which the model will be calibrated. Each column is a security and the rows are the discounted payoffs for each path. This set of instruments includes the payoffs of the smile options (calls and puts) for different maturities, the forward prices at each maturity (the simulation of the paths at each maturity) and the additional constraints presented in section III.

$$\begin{pmatrix} g_{1,1} & g_{1,2} & \cdots & g_{1,N} \\ g_{2,1} & g_{2,2} & & g_{2,N} \\ \vdots & \vdots & \ddots & \vdots \\ g_{v,1} & g_{v,2} & \cdots & g_{v,N} \end{pmatrix} = \begin{pmatrix} \mathbf{g}_1 & \mathbf{g}_2 & \cdots & \mathbf{g}_N \end{pmatrix} \qquad (5)$$

The market prices of these securities are $C_1$ to $C_N$. The whole purpose of the calibration is to find a set of $p_i$ so that equation (6) is satisfied.

$$\sum_{i=1}^{v} p_i g_{i,j} = C_j, \quad j = 1 \; \cdots \; N \qquad (6)$$

Equation (7) shows the formulation of the optimisation problem to get the posterior probabilities $p_i$.

$$\min_{\mathbf{p}} D(\mathbf{p}/\mathbf{q}) = \min_{\mathbf{p}}\left(\ln v + \sum_{i=1}^{v} p_i \ln p_i\right)$$
$$Subject\ to \quad \sum_{i=1}^{v} p_i g_{i,j} = C_j, \quad j = 1 \; \cdots \; N \qquad (7)$$

This optimisation problem can be solved using lagrange multipliers (see [Bertsekas, 1999]) as the min-max problem of equation (8) (same notation as [Avellaneda et al, 2001]).

$$\min_{\lambda}\left(\max_{\mathbf{p}}\left\{-D(\mathbf{p}/\mathbf{q}) + \sum_{j=1}^{N}\lambda_j\left(\sum_{i=1}^{v} p_i g_{i,j} - C_j\right)\right\}\right) \qquad (8)$$

The solution of the maximization problem has been extensively studied by [Cover and Thomas, 1991]. For a given set of $\lambda_j$ the optimal probability is given by equation (9). The normalisation constant Z ensures the probabilities to sum up 1.

$$p_i = \frac{1}{Z(\lambda)}\exp\left(\sum_{j=1}^{N} g_{ij}\lambda_j\right) \text{ with } Z(\lambda) = \sum_{i=1}^{v}\exp\left(\sum_{j=1}^{N} g_{ij}\lambda_j\right) \qquad (9)$$

If these probabilities are replaced in equation (8), the resulting optimisation problem is given by equation (10). Appendix A provides the detail of the calculations.

$$\min_{\lambda} W(\lambda) \text{ where } W(\lambda) = \ln Z(\lambda) - \sum_{j=1}^{N}\lambda_j C_j \qquad (10)$$

At any minimum, the gradient must equal zero. Equation (11) details the calculation of the gradient and shows that the expected values of the discounted payoffs with respect to the posterior probability distribution $\mathbf{p}$ equal the desired option prices $C_k$.

$$\frac{\partial W(\lambda)}{\partial \lambda_k} = \frac{1}{Z(\lambda)}\frac{\partial Z(\lambda)}{\partial \lambda_k} - C_k = 0$$
$$= \frac{1}{Z(\lambda)}\sum_{i=1}^{v} g_{ik}\exp\left(\sum_{j=1}^{N} g_{ij}\lambda_j\right) - C_k = E^{\mathbf{p}}(\mathbf{g}_k) - C_k = 0 \qquad (11)$$

It is shown in [Avellaneda et al, 2001] that when market prices are consistent, this minimum exists and it is unique.

### B. Solution algorithm

This section presents the algorithm to find $\lambda$ such that $W(\lambda)$ is minimized. To achieve a robust, fast and easy to implement algorithm, an iterative method based on the second order Taylor approximation of $W(\lambda)$ around an initial point $\hat{\lambda}$ is used. Equation (12) shows this second order approximation.

$$W(\lambda) \approx W(\hat{\lambda}) + \nabla_\lambda W(\hat{\lambda})(\lambda - \hat{\lambda}) + \frac{1}{2}(\lambda - \hat{\lambda})^T \mathbf{J}(\lambda - \hat{\lambda}) \qquad (12)$$

The matrix $\mathbf{J}$ is the jacobian (second derivatives with respect to $\lambda_i$) and it turns out to be the covariance matrix with respect to the probability $\mathbf{p}$ of the securities $\mathbf{g}_i$ to calibrate (see appendix A). Equation (13) shows a general element $J_{ab}$ of the jacobian and equation (14) shows the gradient and the vector $(\lambda - \hat{\lambda})$.

$$J_{ab} = \frac{\partial W(\lambda)}{\partial \lambda_a \partial \lambda_b} = \mathbf{Cov}^{\mathbf{p}}(\mathbf{g}_a, \mathbf{g}_b) = \sum_{j=1}^{v} g_{aj}g_{bj}p_j - \left(\sum_{j=1}^{v} g_{aj}p_j\right)\left(\sum_{j=1}^{v} g_{bj}p_j\right) \qquad (13)$$

$$\nabla_\lambda W(\hat{\lambda}) = \begin{pmatrix} E^{\mathbf{p}}(\mathbf{g}_1) - C_1 \\ \vdots \\ E^{\mathbf{p}}(\mathbf{g}_N) - C_N \end{pmatrix} \qquad \lambda - \hat{\lambda} = \begin{pmatrix} \lambda_1 - \hat{\lambda}_1 \\ \vdots \\ \lambda_N - \hat{\lambda}_N \end{pmatrix} \qquad (14)$$

The minimization of the second order approximation is a quadratic programming problem which has an analytic solution. Equation (15) shows the optimality conditions of this problem. They are obtained setting to zero the gradient vector of equation (12).

$$\nabla_\lambda W(\hat{\lambda}) + \mathbf{J}(\lambda - \hat{\lambda}) = 0 \qquad (15)$$

The partial solution of the second order approximation is given by equation (16). This partial solution will not be in general the solution that minimizes $W(\lambda)$. However, it provides a point where we can create another second order approximation and start an iteration process which eventually will reach the exact solution. This iteration process will finish when the gradient (11) equals zero. Namely, when the prices of the securities to which the model is calibrated are satisfied.

$$\lambda = \hat{\lambda} - \mathbf{J}^{-1} \nabla_\lambda W(\hat{\lambda}) \qquad (16)$$

As the jacobian is a covariance matrix, it is semi-positive definite and there is always a minimum ($W(\lambda)$ increases with local variations of $\lambda$ around $\hat{\lambda}$). This condition will hold for any point where we do the second order approximation. This means that the optimisation problem is convex and has a minimum.

The steps of the algorithm could be summarised as follows:
1. Simulate $v$ paths at different maturities with a regular Monte Carlo which introduces the particular hypothesis for the underlying process (e.g. stochastic volatility).
2. Calculate the prices $C_k$ of the call and put options of the smile at different maturities using the market volatilities.
3. Calculate the discounted payoffs of all these instruments for each of the simulated paths (the matrix $g_{ij}$).
4. Fix the initial value for the lagrange multipliers $\hat{\lambda}$ equal to zero. According to equation (9), this corresponds to an initial posterior distribution with probabilities equal to each other.
5. Calculate the gradient vector and jacobian according to equations (13) and (14) where the probabilities have previously been calculated with equation (9).
6. Calculate the optimum for the second order approximation using equation (16).
7. Go to step 5 until the gradient is close enough to zero.

There are a few practical tips which should be considered in order to get this algorithm working. When very out of the money smile products are considered, the vector $\mathbf{g}_k$ of discounted payoffs will have zeros in most entries. This means that the covariance of this product with itself or any other products will be very low and the jacobian matrix will be ill conditioned (it will have a row and column of almost zero entries). These products should be removed from the calibration. The products considered in the smile are out of the money calls for strikes over the forward and out of the money puts for strikes below the forward. This allows a symmetric removal of out of the money products (for very low maturities, only a small window of smile options with strikes around the forward will be calibrated). In addition, when there are arbitrage opportunities in the market (e.g. a call and a put with different strikes and same price), the algorithm does not converge. The only way to solve this problem is to remove usually illiquid products which may be mispriced.

The partial solution of step 6 can differ significantly from the solution of the previous step. This is an undesirable effect as the second order approximation of the previous step may no longer be valid. Therefore, a standard under-relaxed Newton update is introduced by a step shortening factor $\alpha$ and the partial solution (17) is replaced by equation (18). This factor is set initially to 0.01 and is multiplied by two after each iteration until $\alpha = 1$.

$$\lambda = \hat{\lambda} - \alpha \mathbf{J}^{-1} \nabla_\lambda W(\hat{\lambda}) \qquad (18)$$

It has been seen that the condition number of the jacobian can increase significantly when the shortening factor is too high. Therefore, the step is divided by 5 when the condition number increases by a factor greater than 10 from one iteration to the following.

### C. Formulation of the problem using weighted least-squares

The formulation in section II.A calibrates exactly to the desired products. When products are mispriced due to low liquidity or wide bid-ask spreads, the algorithm may not converge well and these products should be removed. Therefore, it may be better not to fit the products exactly but to minimize the sum of weighted least-squares of equation (19),

$$\chi_\omega^2 = \frac{1}{2} \sum_{j=1}^{N} \frac{1}{\omega_j} \left( E^\mathbf{p}(\mathbf{g}_j) - C_j \right)^2 \qquad (19)$$

where $\omega_j$ are positive weights. Now, what is minimized is equation (20) without constraints.

$$\min_\mathbf{p} \left[ D(\mathbf{q} / \mathbf{p}) + \chi_\omega^2 \right] \qquad (20)$$

Minimizing (20) with respect to $\mathbf{p}$, is equivalent to minimizing (21) with respect to $\lambda$. The proof of this can be found in [Avellaneda et al, 2001] and will be omitted here. If the equation (21) is compared with equation (10), the difference between exact and weighted least squares fitting is the term after $W(\lambda)$ in (21).

$$\min_\lambda H(\lambda) \quad \text{where} \quad H(\lambda) = W(\lambda) + \frac{1}{2} \sum_{j=1}^{N} \omega_j \lambda_j^2 \qquad (21)$$

Equations (22) and (23) show the minor changes of gradient and jacobian for weighted least squares fitting.

$$\frac{\partial H(\lambda)}{\partial \lambda_a} = E^\mathbf{p}(\mathbf{g}_a) - C_a + \omega_a \lambda_a \qquad (22)$$

$$J_{ab} = \frac{\partial H(\lambda)}{\partial \lambda_a \partial \lambda_b} = \mathbf{Cov}^\mathbf{p}(\mathbf{g}_a, \mathbf{g}_b) + \omega_a \mathbf{1}_{\{a=b\}} \qquad (23)$$

The modified jacobian is the covariance matrix plus a diagonal matrix with the weights of each product. Now it is much better conditioned, as there is no row or column with zero entries. The solution algorithm is the same as section II.B.

### III. Forcing the Martingale Condition

The resulting price process will be feasible for pricing exotic derivatives if it satisfies two conditions: it reproduces the prices of traded derivatives, and it is a martingale. The calibration fulfils the first condition but the second will not in general be satisfied. In particular, given the continuously compounded risk free rate $r_t$ and dividend rate $q_t$ from present time to time $t$, equation (24) shows the process that must be a martingale.

$$S_t e^{-(r_t - q_t)t} \sim \text{Martingale} \qquad (24)$$

Equation (25) shows the martingale condition between two maturities ($t_1$ and $t_2$), where $\mathbf{I}_{t_1}$ represents the information up to time $t_1$.

$$E\left[S_{t_2} e^{-(r_2 - q_2)t_2} / \mathbf{I}_{t_1}\right] = S_{t_1} e^{-(r_1 - q_1)t_1} \qquad (25)$$

To verify this condition within the Monte Carlo environment, it is necessary to consider the paths which go through a window W at $t_1$ as shown in Fig. 1.

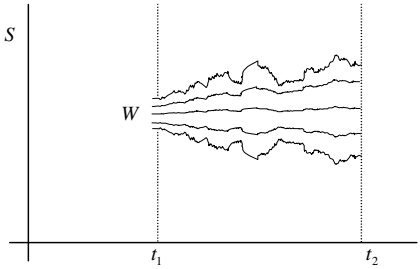

Fig. 1

If small enough windows are considered equation (25) can be understood as equation (26). Equation (27) shows this condition applied to the Monte Carlo paths. Only the paths which go through window W are considered. $S_{j,t_1}$ and $S_{j,t_2}$ are the prices of path $j$ at times $t_1$ and $t_2$ and $F(\cdot)$ is the forward price of the underlying.

$$E\left[S_{t_2} e^{-(r_2 - q_2)t_2} / S_{t_1} \in W \text{ and } \mathbf{I}_{t_1}\right] = E\left[S_{t_1} e^{-(r_1 - q_1)t_1} / S_{t_1} \in W\right] \qquad (26)$$

$$\sum_{\{j/S_{j,t_1} \in W\}} \left(S_{j,t_2} \frac{F(S_{t_1})}{F(S_{t_2})} - S_{j,t_1}\right) \frac{p_j}{\sum_{\{h/S_{h,t_1} \in W\}} p_h} = 0 \qquad (27)$$

Therefore, it is possible to create a synthetic security with zero price and discounted payoffs as shown in equation (28).

$$h_j = \left(S_{j,t_2} \frac{F(S_{t_1})}{F(S_{t_2})} - S_{j,t_1}\right) \mathbf{1}_{\{S_{j,t_1} \in W\}} \quad j = 1 \ \cdots \ v \qquad (28)$$

The narrower the windows are, the more precise the martingale condition is enforced. However, narrower windows imply more simulations and constraints, increasing significantly the computation time. The trade-off chosen is to use the windows defined by the strikes of the options of the smile. Additional windows up and down the highest and lowest strike of the smile are also considered with the similar widths. These additional windows go up to two standard deviations up and down from the current spot level. These standard deviations are considered for the last maturity with ATMF (at-the-money-forward) volatility.

## IV. CASE STUDY

The weighted least squares calibration method has been applied to a geometric cliquet option with payoff (29), where $C$ is a cap set to 1.1, the value date is Jul $21^{st}$ 2005, the dates $t_1$ to $t_6$ are Nov $2^{nd}$ 2005 to 2010 and $t_7$ is Oct $25^{th}$ 2011.

$$\Pi = \max\left(0, \prod_{j=2}^{7}\left[\min\left(\frac{S_{t_j}}{S_{t_{j-1}}}, C\right)\right] - 1\right) \qquad (29)$$

The underlying is the Spanish IBEX index with spot price $S_0 = 10007$. The risk free and dividend rates are respectively 2.95% and 3% (they have been set constant for simplicity), the least squares weights for all constraints are $10^{-7}$ and the volatility surface is presented in Table 1.

TABLE 1: VOLATILITY SURFACE.

| K/Mat | 0.08y | 0.25y | 0.50y | 0.75y | 1y | 2y | 3y | 4y | 5y | 10y |
|---|---|---|---|---|---|---|---|---|---|---|
| 6505 | 35.86 | 30.38 | 25.21 | 24.04 | 22.91 | 22.33 | 22.82 | 23.37 | 23.80 | 24.55 |
| 7005 | 31.86 | 27.38 | 23.26 | 22.39 | 21.41 | 21.33 | 21.77 | 22.37 | 22.90 | 24.00 |
| 7505 | 27.86 | 24.38 | 21.31 | 20.74 | 20.01 | 20.33 | 20.77 | 21.37 | 22.00 | 23.45 |
| 8006 | 23.86 | 21.38 | 19.41 | 19.09 | 18.61 | 19.33 | 19.77 | 20.37 | 21.10 | 22.90 |
| 8506 | 19.86 | 18.38 | 17.51 | 17.44 | 17.31 | 18.33 | 18.77 | 19.47 | 20.30 | 22.35 |
| 9006 | 16.16 | 15.63 | 15.61 | 15.79 | 15.91 | 17.33 | 17.77 | 18.57 | 19.50 | 21.80 |
| 9507 | 12.77 | 13.27 | 13.81 | 14.20 | 14.67 | 16.43 | 16.87 | 17.77 | 18.75 | 21.30 |
| 10007 | 10.42 | 11.37 | 12.21 | 12.89 | 13.53 | 15.53 | 15.97 | 16.97 | 18.05 | 20.80 |
| 10507 | 10.02 | 10.31 | 11.13 | 11.79 | 12.46 | 14.68 | 15.17 | 16.17 | 17.35 | 20.30 |
| 11008 | 11.00 | 10.03 | 10.65 | 10.97 | 11.71 | 13.88 | 14.47 | 15.47 | 16.70 | 19.85 |
| 11508 | 13.00 | 9.98 | 10.50 | 10.47 | 11.21 | 13.28 | 13.87 | 14.77 | 16.05 | 19.45 |
| 12008 | 15.50 | 10.08 | 10.40 | 10.07 | 10.71 | 12.68 | 13.27 | 14.17 | 15.45 | 19.10 |
| 12509 | 18.50 | 10.48 | 10.40 | 9.87 | 10.31 | 12.18 | 12.87 | 13.77 | 14.85 | 18.80 |
| 13009 | 21.50 | 10.98 | 10.50 | 9.77 | 9.91 | 11.68 | 12.47 | 13.37 | 14.25 | 18.50 |
| 13509 | 24.50 | 11.58 | 10.80 | 9.67 | 9.61 | 11.38 | 12.07 | 12.97 | 13.85 | 18.35 |

The paths were generated with a regular Monte Carlo with at-the-money-forward (ATMF) deterministic volatility (the volatility level is interpolated in Table 1 using the forward at each maturity). The smile has been fitted to the seven maturities of the cliquet option and the martingale condition has been imposed on windows starting on 0.35 times the spot up to 2.25 at each pair of consecutive maturities. The width is the distance between consecutive strikes of matrix in Table 1. Windows outside this matrix use the last upper and lower window width. From 346 constraints ($15 \cdot 7 = 105$ smile options, $39 \cdot 6 = 234$ martingale conditions and 7 forwards) 131 were removed (the 5 upper and 4 lower options of the smile at the first maturity, the upper smile option of the second maturity and the rest were out of the money martingale conditions). All option prices were fitted with a precision better than $10^{-5}$ (errors less than a tenth of a basis point). The total number of paths is 20000. The algorithm converges after 14 iterations.

TABLE 2: PRICE RESULTS

| ATMF | SML $t_1$ to $t_6$ | SML $t_1$ to $t_6$ Mtgl |
|---|---|---|
| 0.0332 | 0.0387 | 0.0547 |

Table 2 compares three prices. The first one corresponds to the regular Monte Carlo with ATMF volatility (all paths have the same probability). The last two prices correspond to the weighted Monte Carlo fitting the smile at every maturity without imposing the martingale condition and imposing it. All prices are calculated with the same paths only changing the path probabilities according to the calibration. Enforcing the martingale condition makes a big difference: the option is 160bp (basis points) more expensive. This is in agreement with trader experience and market pricing. Fig. 2 presents the distribution of the underlying at $t_4$ and shows how the distribution is skewed to lower values (the left queue is considerably bigger than the right one).

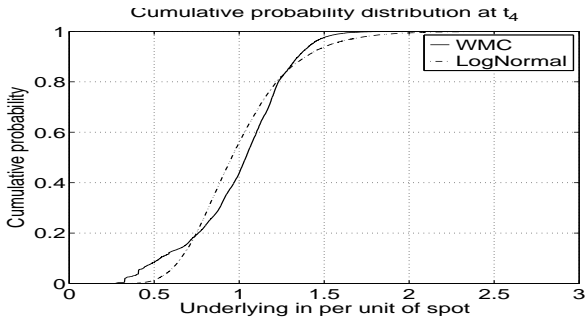

Fig. 2: Cumulative probability distribution at $t_4$.

To explain the difference in price, consider Fig. 3 with the martingale condition mismatches with martingale constraints (right plot) and without them (left plot). They correspond to the period $t_4$ to $t_5$ (the rest of the periods look very similar). The horizontal axis has the levels of the martingale windows and the vertical axis shows equation (27) both in per unit of the spot. The dotted line corresponds to the regular Monte Carlo (RMC) and the solid line to the weighted Monte Carlo (WMC). Both plots show that the RMC reasonably satisfies the martingale condition (the mismatches are below 2%). The left plot shows that the WMC without martingale constraints does not satisfy the condition. The underlying process is a super-martingale at $t_4$ for levels between 0.7 to 1.2 of the spot and a sub-martingale outside them. The right plot clearly satisfies the condition because it is enforced.

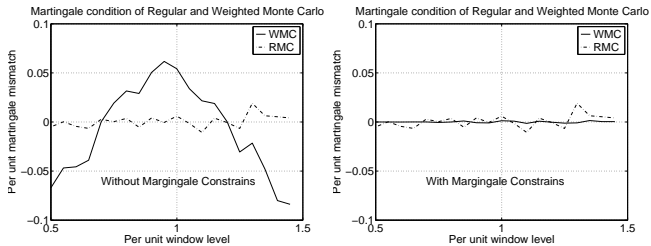

Fig. 3: Martingale condition mismatches from $t_4$ to $t_5$ without forcing the condition (left) and forcing it (right).

At first sight it seems reasonable to think that the option should be more expensive for the left plot of Fig. 3, as more likely underlying values around the spot would have a higher expected value (higher return). To prove that this perception is false, Fig. 4 shows the cumulative probability distributions of non-paying (left) and paying (right) paths at $t_7$ when the martingale condition is (WMC MTGL) and is not (WMC) enforced. A paying path is a path for which equation (29) is different from zero.

These distributions are not conditional and therefore about 70% of the paths do not contribute to the payoff whereas about 30% indeed do. The right plot of Fig. 4 shows that all the paying paths of the weighted Monte Carlo where the martingale condition is enforced (solid line) have higher probabilities than when it is not (dotted line). That is why the price is significantly higher. The left plot of Fig. 4 shows that the paths which finish below 1.2 times the spot have the same probabilities with and without martingale condition. These paths correspond to decreasing paths which do not contribute to the payoff. The paths which finish above 1.2 times the spot change significantly their probabilities with and without the martingale condition. Fig. 3 (left) shows that paths whose spot is between 0.7 and 1.2 are more likely to go up and above 1.2 are more likely to go down. Therefore, it is clear from the martingale mismatches of Fig. 3 (left) that zigzagging paths are over weighted (otherwise the plot would be flat). These zigzagging paths do not contribute significantly to the payoff (very negative returns decrease the payoff and very positive returns do not compensate the payoff as they are capped). Therefore, the price of the option reduces. When the martingale condition is enforced, these zigzagging paths reduce their probabilities as they are responsible for the distortion of the martingale condition and allow other paying paths to increase their probabilities, rising the price of the option.

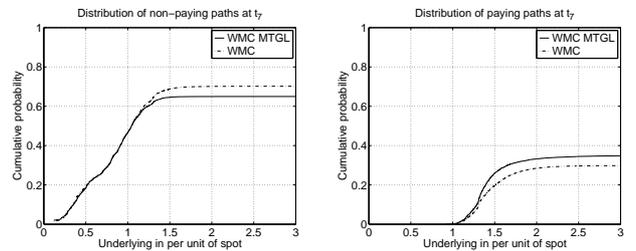

Fig. 4: Cumulative unconditioned probability distribution of non-paying (left) and paying paths (right).

## V. CONCLUSIONS

When the weighted Monte Carlo is calibrated to several maturities with steep smiles, the resulting underlying process after calibration may not be a martingale. For certain type of path dependent options this may have a big impact in price.

A simple solution to overcome this problem is presented. It consists of adding additional constraints to the problem.

A new robust, fast and easy-to-implement calibration algorithm is proposed. It can cope with a significant number of constraints.

The improved method is applied to a well-known geometric cliquet option with a big impact in price also in agreement with trader and market experience.

## VI. APPENDIX A

### A. Derivation of the optimisation problem

This section derives the final optimisation problem of equation (10) when the probabilities of equation (9) are replaced in (8).

$$\min_{\boldsymbol{\lambda}}\left( \max_{\mathbf{p}}\left\{ -D(\mathbf{p}/\mathbf{q}) + \sum_{j=1}^{N} \lambda_j \left( \sum_{i=1}^{\nu} p_i g_{i,j} - C_j \right) \right\} \right)$$

$$= \min_{\boldsymbol{\lambda}}\left( -\ln \nu - \sum_{i=1}^{\nu} p_i \ln p_i + \sum_{j=1}^{N} \lambda_j \left( \sum_{i=1}^{\nu} p_i g_{i,j} - C_j \right) \right)$$

$$= \min_{\lambda} \left( -\sum_{i=1}^{\nu} p_i \left( -\ln Z(\lambda) + \sum_{i=1}^{\nu} p_i g_{i,j} \right) + \sum_{j=1}^{N} \lambda_j \sum_{i=1}^{\nu} p_i g_{i,j} - \sum_{j=1}^{N} \lambda_j C_j \right)$$

$$= \min_{\lambda} \left( \ln Z(\lambda) \sum_{i=1}^{\nu} p_i - \sum_{i=1}^{\nu} p_i \sum_{i=1}^{\nu} g_{i,j} \lambda_j + \sum_{j=1}^{N} \lambda_j \sum_{i=1}^{\nu} p_i g_{i,j} - \sum_{j=1}^{N} \lambda_j C_j \right)$$

$$= \min_{\lambda} \left( \ln Z(\lambda) \sum_{i=1}^{\nu} p_i - \sum_{i=1}^{\nu} \sum_{j=1}^{N} p_i g_{i,j} \lambda_j + \sum_{j=1}^{N} \sum_{i=1}^{\nu} p_i g_{i,j} \lambda_j - \sum_{j=1}^{N} \lambda_j C_j \right)$$

$$= \min_{\lambda} \left( \ln Z(\lambda) - \sum_{j=1}^{N} \lambda_j C_j \right)$$

$$= \min_{\lambda} W(\lambda) \quad \text{where} \quad W(\lambda) = \ln Z(\lambda) - \sum_{j=1}^{N} \lambda_j C_j \quad (30)$$

### B. Derivation of the Jacobian

The Jacobian is the matrix of second derivatives. Equation (31) shows the element at the row $u$ and column $v$ of this matrix.

$$J_{uv} = \frac{\partial W(\lambda)}{\partial \lambda_u \partial \lambda_v} = \frac{\partial}{\partial \lambda_v} \left( E^p(g_u) - C_u \right) = \frac{\partial E^p(g_u)}{\partial \lambda_v} = Cov^p(g_u, g_v) \quad (31)$$

The derivative of $W(\lambda)$ with respect to $\lambda_u$ is given by equation (11) and equation (32) presents the derivative of the expected value of a general payoff vector $\mathbf{h}$.

$$\frac{\partial E^p(\mathbf{h})}{\partial \lambda_u} = \frac{\partial}{\partial \lambda_u} \left( \sum_{i=1}^{\nu} p_i h_i \right)$$

$$= \frac{\partial}{\partial \lambda_u} \left( \frac{1}{Z(\lambda)} \sum_{i=1}^{\nu} \exp \left( \sum_{j=1}^{N} g_{ij} \lambda_j \right) h_i \right)$$

$$= \frac{-1}{Z(\lambda)^2} \frac{\partial Z(\lambda)}{\partial \lambda_u} \sum_{i=1}^{\nu} \exp \left( \sum_{j=1}^{N} g_{ij} \lambda_j \right) h_i + \frac{1}{Z(\lambda)} \sum_{i=1}^{\nu} \exp \left( \sum_{j=1}^{N} g_{ij} \lambda_j \right) h_i g_{iu}$$

$$= \frac{-1}{Z(\lambda)} \frac{\partial Z(\lambda)}{\partial \lambda_u} \sum_{i=1}^{\nu} p_i h_i + \sum_{i=1}^{\nu} p_i h_i g_{iu}$$

$$= \sum_{h=1}^{\nu} p_h g_{hu} \sum_{i=1}^{\nu} p_i h_i + \sum_{i=1}^{\nu} p_i h_i g_{iu}$$

$$= E^p(g_u) E^p(\mathbf{h}) - E^p(g_u \mathbf{h}) = Cov^p(g_u, \mathbf{h}) \quad (32)$$

The authors want to thank R. Ordovás and other member of the area for their helpful comments.

**A. Elices** earned a PhD in Power Systems engineering at Pontificia Comillas University (Madrid, Spain) and a Masters in Financial Mathematics in the University of Chicago. He is a senior quant team member in the Model Validation group of the Risk Department at Santander in Madrid after working in a hedge fund in New York.

**E. Giménez** earned a Masters in Artificial Intelligence at Institut d'Investigació en Intel.ligència Artificial –(IIIA-CSIC, Barcelona, Spain) and a Masters in Financial Mathematics at Grupo Analistas (Madrid, Spain). He is a senior quant team member in the Model Development Group of Caixabank after working in the Model Validation Group of Santander and in the consulting division at Grupo Analistas.